\documentclass[twocolumn]{aastex62}

\usepackage{amsmath}
\usepackage{amssymb}
\usepackage{verbatim}
\usepackage{subfigure}
\usepackage{hyperref}

\graphicspath{{figs/}}

\newcommand{\E}[1]{\ensuremath{\times 10^{#1}}}

\newcommand{\diff}{\mathop{}\!\mathrm{d}}

\newcommand{\like}{\ensuremath{\mathcal{L}}}
\newcommand{\normal}{\ensuremath{\mathcal{N}}}

\newcommand{\rt}{\ensuremath{r_\mathrm{tidal}}}
\newcommand{\sersic}{S\'{e}rsic}
\newcommand{\Reff}{\ensuremath{R_\mathrm{e}}}

\newcommand{\kms}{\ \ensuremath{\mathrm{km \ s}^{-1}}}
\newcommand{\vir}{\ensuremath{\mathrm{200c}}}

\shorttitle{Jeans modeling of NGC 1052-DF2}
\shortauthors{Wasserman et al.}

\begin{document}

\title{A Deficit of Dark Matter from Jeans Modeling of the Ultra-diffuse Galaxy NGC~1052-DF2}

\correspondingauthor{Asher Wasserman}
\email{adwasser@ucsc.edu}

\author[0000-0003-4235-3595]{Asher Wasserman}
\affil{Department of Astronomy \& Astrophysics, University of California-Santa Cruz, Santa Cruz, CA 95064, USA}
\author[0000-0003-2473-0369]{Aaron J. Romanowsky}
\affil{Department of Physics \& Astronomy, San Jos\'{e} State University, One Washington Square, San Jose, CA 95192, USA}
\affil{University of California Observatories, 1156 High Street, Santa Cruz, CA 95064, USA}
\author{Jean Brodie}
\affil{Department of Astronomy \& Astrophysics, University of California-Santa Cruz, Santa Cruz, CA 95064, USA}
\affil{University of California Observatories, 1156 High Street, Santa Cruz, CA 95064, USA}
\author{Pieter van Dokkum}
\affil{Astronomy Department, Yale University, New Haven, CT 06511, USA}
\author{Charlie Conroy}
\affil{Department of Astronomy, Harvard University, Cambridge, MA 02138, USA}
\author{Roberto Abraham}
\affil{Department of Astronomy \& Astrophysics, University of Toronto, 50 St. George Street, Toronto, ON M5S 3H4, Canada}
\author{Yotam Cohen}
\affil{Astronomy Department, Yale University, New Haven, CT 06511, USA}
\author{Shany Danieli}
\affil{Astronomy Department, Yale University, New Haven, CT 06511, USA}

\begin{abstract}

 The discovery of the ultra-diffuse galaxy NGC~1052-DF2 and its peculiar population of star clusters has raised new questions about the connections between galaxies and dark matter halos at the extremes of galaxy formation.  In light of debates over the measured velocity dispersion of its star clusters and the associated mass estimate, we constrain mass models of DF2 using its observed kinematics 
with a range of priors on the halo mass. Models in which the galaxy obeys a standard stellar-halo mass relation are in tension with the data and also require a large central density core. Better fits are obtained when the halo mass is left free, even after accounting for increased model complexity. The dynamical mass-to-light ratio for our model with a weak prior on the halo mass is $1.7^{+0.7}_{-0.5} \ M_\odot / L_{\odot, V}$, consistent with the stellar population estimate for DF2. We use tidal analysis to find that the low-mass models are consistent with the undisturbed isophotes of DF2. Finally we compare with Local Group dwarf galaxies and demonstrate that DF2 is an outlier in both its spatial extent and its relative dark matter deficit.

\end{abstract}

\keywords{galaxies: halos --- galaxies: individual (NGC 1052, DF2) --- galaxies: kinematics and dynamics}

\section{Introduction}\label{sec:intro}

Ultra-diffuse galaxies (UDGs) were recently recognized as a ubiquitous class of 
low-surface-brightness stellar systems with
luminosities like dwarf galaxies but sizes like giants \citep{Dokkum2015,Yagi2016}.
They are found in all environments from clusters and groups to the field 
(e.g., \citealt{Martinez-Delgado2016,Burg2017}), 
and appear to originate from multiple formation channels, including 
an extension of normal dwarfs to lower surface brightness,
as tidal debris, and
perhaps as ``failed'' galaxies
(e.g., \citealt{Peng2016,Greco2018,Pandya2018}).

The failed-galaxy scenario was motivated partly by inferences of UDG halo masses based on
dynamics and on number-counts of globular star-clusters -- 
masses that in some cases appear significantly
higher than for the overall dwarf-galaxy population
\citep{Beasley2016,Dokkum2016,Dokkum2017,Amorisco2018,Lim2018}.
The implication is that the stellar-to-halo mass relation 
\citep[SHMR; e.g.,][]{Moster2013, Rodriguez-Puebla2017} 
for luminous dwarf galaxies ($L \sim 10^8 L_\odot$) may have a much larger scatter than was
presumed, requiring revisions in galaxy formation models at halo masses of $\sim 10^{11} M_\odot$
(see also \citealt{Smercina2018a}).

In this context, one of the nearest known UDGs, NGC~1052-DF2 in a galaxy group at $\sim$~20~Mpc
(\citealt{Fosbury1978}, \citealt{Karachentsev2000}; \citealt[hereafter vD+18a]{Dokkum2018}),
presents a valuable opportunity for detailed dynamical study.
vD+18a used deep Keck spectroscopy to measure radial velocities for 10 luminous star-clusters
around DF2, 
estimating its dynamical mass within a radius of $\sim$~8~kpc
(cf.\ Virgo-UDGs work by \citealt{Beasley2016,Toloba2018}).
The result was very surprising: rather than an unusually {\it high} mass-to-light ratio ($M/L$) as
found for previous UDGs, the $M/L$ was unusually {\it low}, and consistent with harboring 
no dark matter (DM) at all.

The low/no-DM result generated spirited debate, much of which focused on how best to estimate the
intrinsic velocity dispersion $\sigma$ of DF2 (e.g., \citealt{Martin2018a,Laporte2018,Dokkum2018b}).

However, the more fundamental question is what range of halo mass profiles is permitted by the data,
which we examine in detail in this Letter. We adopt a generative modeling approach where the individual velocity measurements are mapped statistically onto halo parameter space, without the intervening steps of estimating $\sigma$ and applying a mass estimator.  In addition to deriving constraints on the dynamical mass profile, we consider the potential impact of tidal stripping, and furthermore compare DF2 with Local Group dwarfs.

\section{Observational Constraints}\label{sec:data}

NGC~1052-DF2 has position, redshift, surface brightness fluctuation (SBF),
and tip of the red giant branch measurements
all consistent with being a satellite of the giant elliptical galaxy NGC~1052 (vD+18a; \citealt{vD+2018d}).  We adopt a distance of 19~Mpc, matching the measured SBF distance to DF2 \citep{Cohen2018}, while allowing for a $\pm 1$~Mpc uncertainty in our analysis\footnote{A distance of 13 Mpc has been proposed \citep{Trujillo2018a}, but see \cite{vD+2018d} for an in-depth discussion of the evidence for the greater distance.}.

The UDG surface brightness follows a \sersic{} profile with index $n = 0.6$, effective radius $\Reff{} = 22.6\arcsec$ (2.08 kpc), and 
total luminosity of $1.2\E{8} \ L_{V,\odot}$.
For the stellar $M/L$, 
we adopt a Gaussian prior with mean of $\Upsilon_{*, V} = 1.7$ in Solar units
and standard deviation of 0.5
(based on stellar population modeling; vD+18a; \citealt{Dokkum2018a}).
We truncate this distribution to be between 0.1 and 10.

NGC~1052-DF2 has ten star-clusters with radial velocity measurements in vD+18a.
We use one updated velocity from \cite{Dokkum2018b}; this has only a mild impact on the results.
Although the mass uncertainties from using so few tracers is relatively large (as we will find here), there is ample precedent in the literature for drawing meaningful conclusions from small sample sizes
(\citealt{Aaronson1983,Kleyna2005,Chapman2005,Brown2007,Koposov2015}).

The surface-density distribution of the star-cluster population is highly unconstrained.  We assume an exponential 
distribution of tracers (i.e., a \sersic{} profile with $n = 1$) 
where the half-number radius is drawn from a Gaussian prior with a mean of the observed half-number radius (32\arcsec) and standard deviation of 10\arcsec.  We truncate this distribution to be between 10\arcsec\ and 70\arcsec.  Our adopted mean half-number radius is 40\% larger than $R_\mathrm{e}$
of the galaxy diffuse starlight, 
consistent with studies of the star-cluster systems of other UDGs (\citealt{Peng2016,Dokkum2017,Toloba2018}; cf.\ \citealt{Forbes2017}).

\section{Jeans Modeling Methods}\label{sec:methods}

We use the Bayesian Jeans modeling formalism of \citeauthor{Wasserman2018} (2018, in press) to infer the mass distribution of DF2.
Here a given mass profile and a tracer density profile are linked to 
a predicted line-of-sight velocity dispersion profile $\sigma_{\rm J}(R)$.
The assumptions include spherical symmetry, dynamical equilibrium, and velocity-dispersion anisotropy
($\beta = 1 -  	\sigma_t^2 / \sigma_r^2$) 
that is constant with galactocentric radius.
(There is no evidence for rotation in the system, although individual velocity uncertainties are too large for strong constraints; vD+18a).
We adopt a Gaussian prior on $\tilde{\beta} = - \log_{10}(1 - \beta)$ with a mean of 0 (isotropic) and standard deviation of 0.5 (truncated to the range of $\tilde{\beta} = -1$ to +1).

Since we do not directly constrain the dynamical mass beyond $\sim 8$~kpc, we must rely on priors on the halo characteristics -- on the DM profile shape, and
also on expected correlations between halo mass, concentration, and stellar mass.

We model the mass distribution as the sum of the stellar mass, with spatial distribution described in Section~\ref{sec:data}, and a DM halo.  
For the halo density distribution we use the generalized Navarro--Frenk--White (gNFW) profile,
\begin{equation}\label{eq:gnfw}
	\rho(r) = \rho_s \left(\frac{r}{r_s}\right)^{-\gamma} \left(1 + \frac{r}{r_s}\right)^{\gamma - 3}
\end{equation}
where $r_s$ is the scale radius, $\rho_s$ is the scale density, and $\gamma$ quantifies the inner log-slope.  For $\gamma = 1$, this matches the usual NFW halo model \citep{Navarro1997}, but letting $\gamma$ vary below 1 allows for models which have a cored, shallower density 
profile.%\footnote{We have also tried a model that allows for a separate, smaller core radius ($r_{\rm c} \leq r_s)$, which may be more physically plausible.  The results were similar but with somewhat more central DM required in the standard model.}.

We re-parameterize the halo in terms of virial mass ($M_\vir$) and concentration ($c_\vir$), where
\begin{equation}\label{eq:virial}
	M_\vir = 200\rho_\mathrm{crit} \frac{4\pi r_\vir^3}{3}
\end{equation}
and $c_\vir = r_\vir / r_s$.

We then consider two flavors of mass models: 
one in which the stellar and halo masses are drawn from a SHMR,
and one where the stellar and dark masses are decoupled.  
For the latter model, we use a uniform prior on $\log_{10} M_\vir / M_\odot$ between 2 and 15.
 This effectively allows for the case of no DM, since the stellar mass is $\log_{10} M_\star / M_\odot \sim 8.3$.  

For both types of models we assume that the halo concentration is drawn from a mass--concentration relation (MCR; \citealt{Diemer2015,Diemer2017}) based on the \emph{Planck 2015} cosmology.
We use a log-normal distribution about this expected concentration with a scatter of $0.16$ dex. 

For the SHMR we use the $z = 0$ relation of \cite{Rodriguez-Puebla2017},
where halos with mass $M_\vir \sim 10^{10.8} M_\odot$ host galaxies with $M_*$ similar to DF2
(note that for a satellite galaxy such as DF2, the halo mass is pre-infall, before tidal stripping).
We allow for variation around this mean relation through a variable scatter:
\begin{equation}
	\label{eq:scatter}
    \sigma_{\log M_*} = 0.2 - 0.26 (\log M_\mathrm{vir} - \log M_1)
\end{equation}
below virial masses of $M_1 = 10^{11.5} \ M_\odot$ 
(note $M_\mathrm{vir} \neq M_\vir$; at $M_1$, $M_\vir \sim 0.9 M_\mathrm{vir}$), 
while at higher masses, $\sigma_{\log M_*}$ is 
a constant $0.2$ dex scatter \citep{Garrison-Kimmel2017, Munshi2017}.

Given the wide range of possible 
baryonic effects on the inner slope of DM halos
\citep{Oh2011, Adams2014, Pineda2017}, 
we adopt a uniform prior on $\gamma$
between 0 and 2.

To connect the Jeans model predictions for $\sigma_{\rm J}$ to the velocity observations,
we use a Gaussian likelihood for the probability of drawing data, $v_i$, 
given the location $R_i$ and the various model parameters $\theta$,

\begin{align}\label{eq:likelihood}
	\like(v_i | R, \theta) &= \normal\left(v_i - v_\mathrm{sys}, \ \sigma^2 = \sigma_{\rm J}^2(R | \theta) + \delta v_i^2\right) \\
    &= \left(2\pi\sigma^2\right)^{-1/2} \exp\left(-\frac{(v_i - v_\mathrm{sys})^2}{\sigma^2}\right) \nonumber
    \end{align}
where $v_\mathrm{sys}$ is the systemic velocity
(drawn from a Gaussian prior with a mean of the observed velocities, $1801.6$\kms, and with a 5\kms\ standard deviation), 
and $\delta v_i$ is the measurement uncertainty.  

We draw from our posterior with the \emph{emcee} Markov Chain Monte Carlo (MCMC) ensemble sampler \citep{Foreman-Mackey2013}. We run our sampler with 128 walkers for 2000 iterations, rejecting the first 1500 to ensure fully-mixed chains.  The posterior distributions of $v_\mathrm{sys}$, $\Upsilon_{*}$, and distance closely match those of the associated prior distributions. For the inference with the SHMR-informed prior, the posterior distribution of the star-cluster system $R_\mathrm{e}$ is slightly lower (with median of $26\arcsec$).  The weak-prior model prefers a slightly tangential orbital anisotropy, although consistent with isotropy, while the posterior anisotropy in the SHMR-prior model matches the prior.

\section{Halo Mass Inferences}\label{sec:inference}

\begin{figure*}
	\includegraphics[width=\columnwidth]{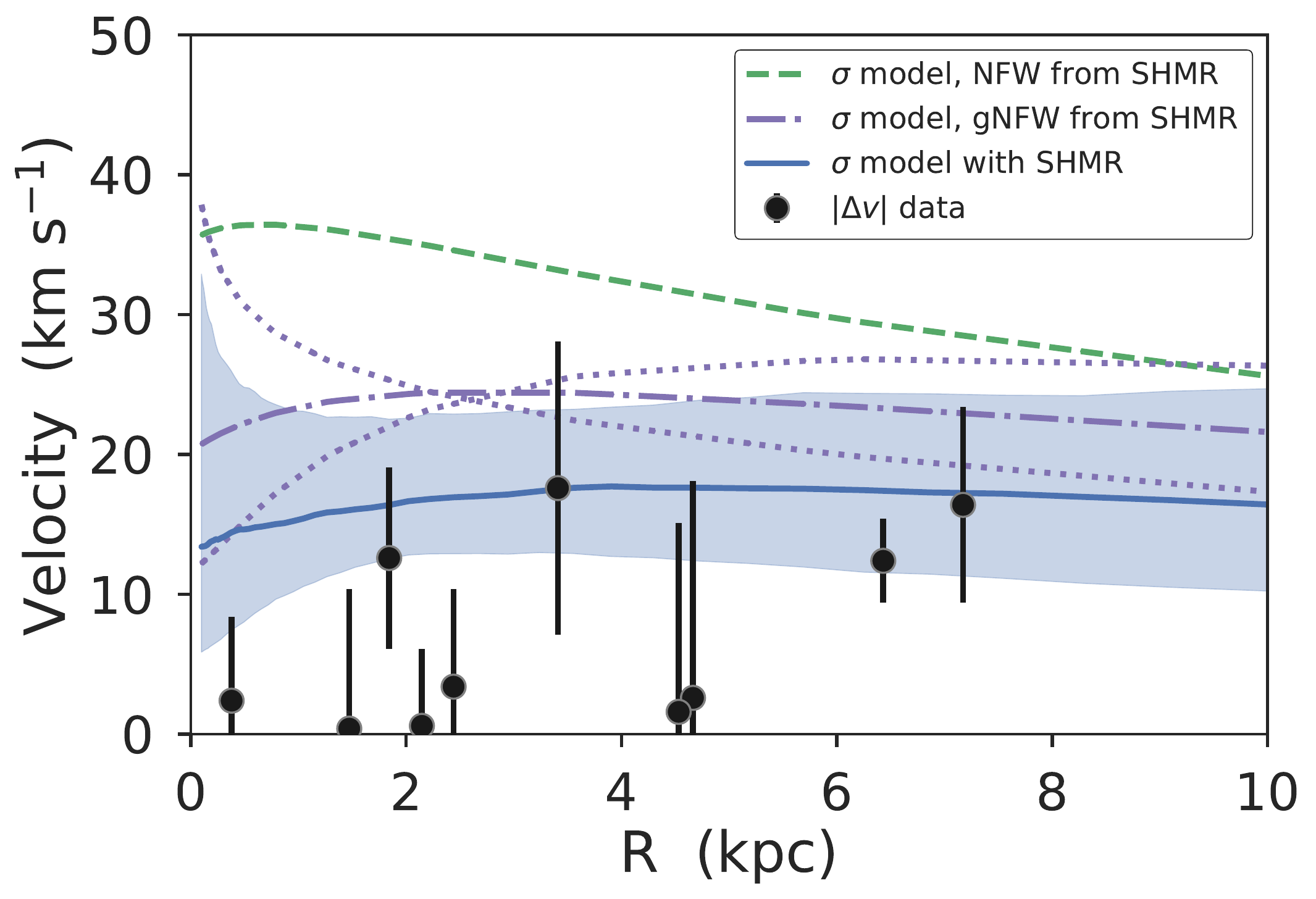}
    \includegraphics[width=\columnwidth]{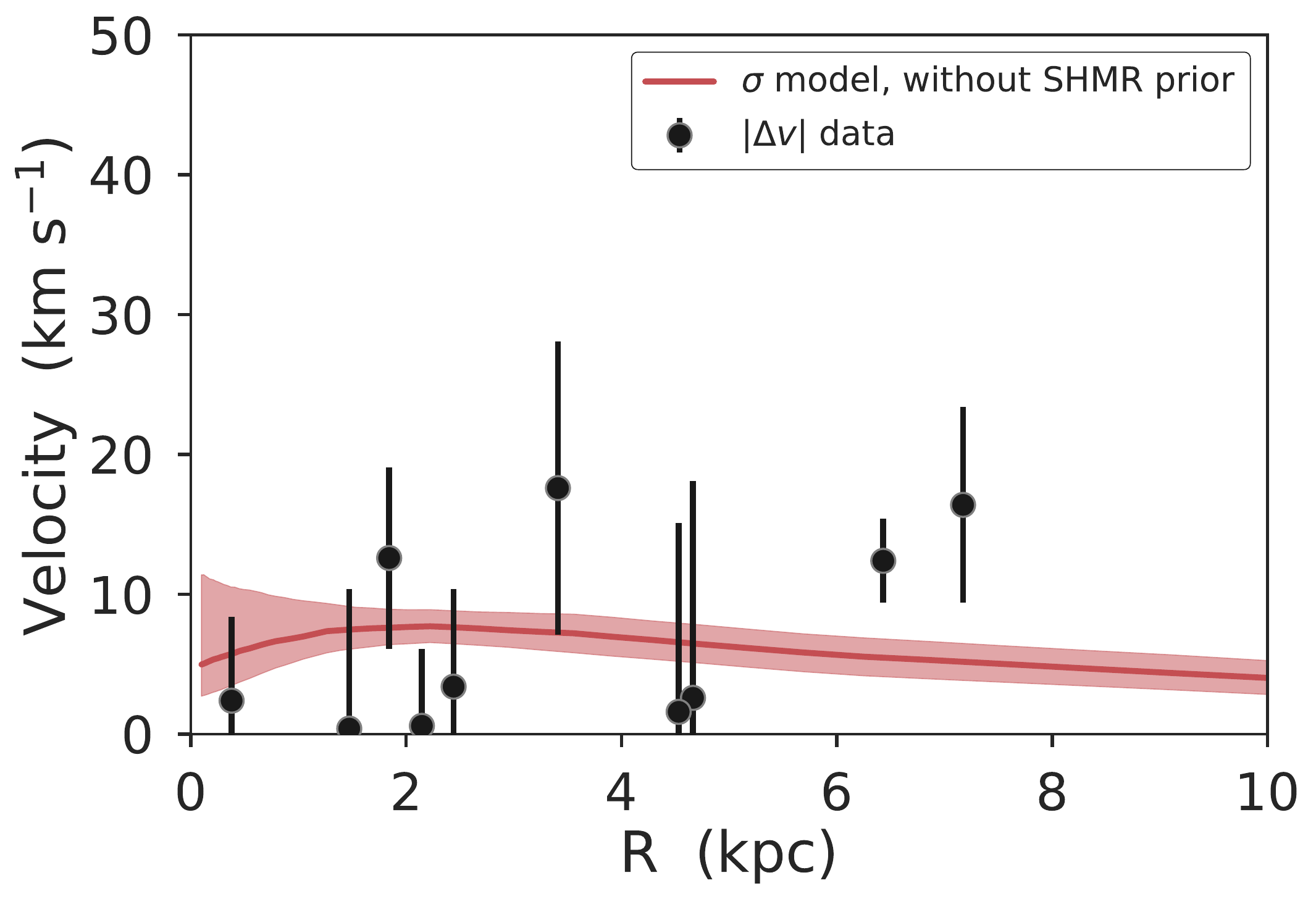}
    \caption{DF2 observed star-cluster velocity offsets (points with error-bars), compared 
    with the posterior predictive distribution of the velocity dispersion profiles
    associated with the star+halo model fit with freely-varying anisotropy and \Reff{}.
    The shaded regions give the inner 68\% of samples. 
    \emph{Left: } The dashed green curve shows an isotropic model with a standard DM halo ($\gamma=1$ cusp) and halo mass fixed to the SHMR mean.  The dot-dashed purple curve is for a cored DM halo ($\gamma = 0.2$), with fixed halo mass, and isotropic orbits.  The dotted purple lines around this curve show the effect of assuming radial (falling profile) and tangential (rising profile) anisotropy.
    The blue solid curve shows a cored halo with mass informed by a log-normal prior about a standard SHMR.
    \emph{Right: } The red solid curve shows the model fit with the relaxed prior on halo mass -- a model that we see is less in tension with the data than the models with large amounts of DM (left panel).
	}
    \label{fig:postPred}
\end{figure*}

Before discussing the best-fitting results,
in Figure~\ref{fig:postPred}
we present a comparison between the data and a simple model
with a cuspy NFW halo that follows the mean SHMR, assuming isotropic orbits.
The individual star-cluster velocity measurements (absolute value relative to $v_{\rm sys}$)
versus galactocentric radius 
are shown along with a model line-of-sight $\sigma$ profile 
(dashed-green curve).
It is clear that this is not a favorable model: $\sim 3$ of the 
observed velocities should lie above the curve, 
which has a spatially-averaged $\sigma \sim 36$\kms, compared to an observed $\sigma \sim 5$--$10$\kms.	

%\deleted{and increases toward the center -- compared to an observed $\sigma \sim $~5--10\kms\ that if anything {\it decreases} toward the center.  This model has radial anisotropy ($\beta \sim 0.5$) and a low star-cluster $R_\mathrm{e} \sim 20\arcsec$ (which lies near the prior boundary).}

\begin{figure*}
	\includegraphics[width=\columnwidth]{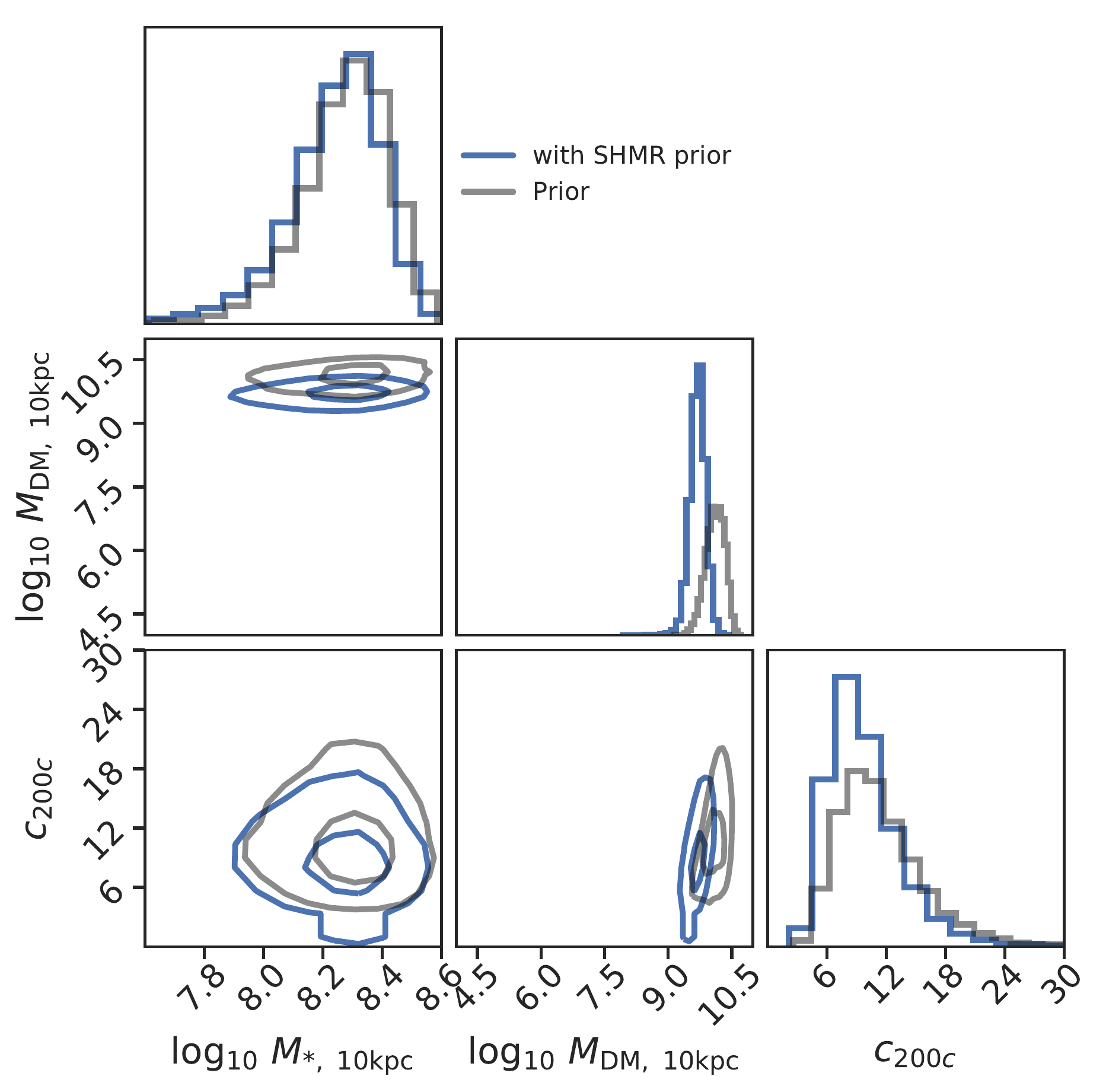}
	\includegraphics[width=\columnwidth]{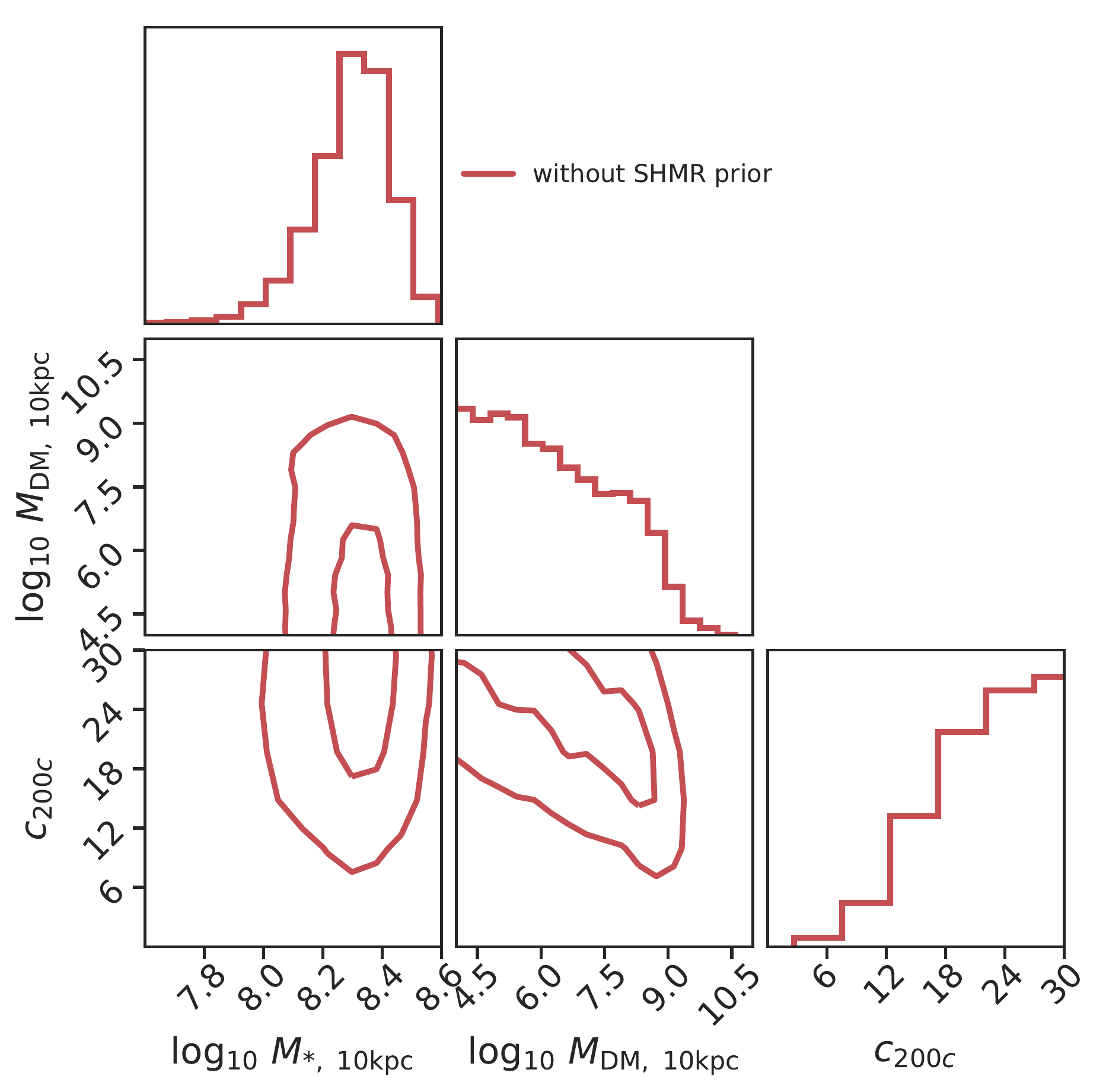}
    \caption{Distributions of select model parameters. The contours showing the covariance between the two parameters are placed at 1- and 2-$\sigma$ intervals.  Masses are in $M_\odot$.
    \emph{Left:} For the model with the SHMR prior (in blue).  The prior distribution is shown in gray. From left to right, the parameters are the stellar mass within 10 kpc, the DM mass within 10 kpc, and the halo concentration.  
    \emph{Right:} The same model parameters but for the model without the SHMR prior.  
	We see that the SHMR prior model largely recovers the prior distribution, though with slightly lower halo mass, while the data-driven model has a halo mass that hits the prior lower-boundary.}
    \label{fig:mass_post}
\end{figure*}

This is not however the only plausible model, as there is
scatter in the predicted SHMR and in the halo concentration. 
Furthermore, UDGs and luminous dwarfs in general are expected to 
inhabit cored DM halos (\citealt{Chan2015,DiCintio2017}).
Allowing for a DM core (dot-dashed purple curve) reduces the tension with the data somewhat ($\sigma \sim 22$\kms).  
Introducing scatter in the SHMR and the MCR as discussed in Section~\ref{sec:methods}, we present the best ``standard'' model from our MCMC fitting, including a freely varying orbital anisotropy,
as a solid-blue curve with uncertainty envelope in Figure~\ref{fig:postPred}.
This model dispersion profile is fairly constant
with spatially-averaged $\sigma = 17^{+6}_{-4}$\kms\
and appears more reasonably close to the data, although still in tension with
the many observed near-zero relative velocities.
The posteriors on some key model parameters are:
$\beta = 0.0^{+0.7}_{-2.5},$
$\log_{10} M_\vir / M_\odot = 10.7^{+0.2}_{-0.3}$, $c_\vir = 9^{+4}_{-3}$
(implying $r_s = 8^{+4}_{-3}$~kpc), 
and $\gamma = 0.2^{+0.3}_{-0.2}$, although we note that the samples of $\gamma$ hit the prior boundary at 0. 
This is a model solution with a normal halo and concentration
(consistent with the priors: see Figure~\ref{fig:mass_post}, {\it left}) but a large central density core -- 
strongly disfavoring the NFW model.

We next consider a model that allows for deviation from the 
standard SHMR, along with a free central DM slope,
while still imposing the standard prior on halo mass versus concentration.
We find that the DM halo all but disappears,
with $M_{\rm 200c} < 1.2\E{8} \ M_\odot$ 
$(M_{\rm DM}/M_* < 0.6)$ 
at the 90$^{\rm th}$ percentile.
The posterior velocity dispersion profile is shown in 
Figure~\ref{fig:postPred} (right), with an average
$\sigma = 7\pm1$\kms.  This model prefers a more tangential $\beta = -1.0^{+1.2}_{-2.7}$.

For a measure of relative predictive accuracy of these two models, we use the Watanabe--Akaike Information Criterion (WAIC), an approximation of cross-validation 
\citep{Gelman2013}, defined as
\begin{align}
	\mathrm{WAIC} =& -2\sum_i^n \ln \int \like(v_i \ | \ \theta) p_\mathrm{post}(\theta) \diff\theta \\
    & +4\sum_i^n \mathrm{var}_\mathrm{post}\left[\ln\like(v_i \ | \ \theta)\right] \nonumber
\end{align}
where $p_\mathrm{post}(\theta)$ is the posterior distribution, $\like(v_i | \theta)$ is the likelihood, and $\mathrm{var}_\mathrm{post}$ is the variance over the posterior.  The first term measures the predictive accuracy marginalized over the posterior distribution, while the second term penalizes for model complexity by computing an approximation of the effective number of model parameters (analogous to reduced $\chi^2$).  We find $\Delta\mathrm{WAIC} = 1.5$ (equivalent to a model odds ratio of $\sim 2$), slightly favoring the model without the SHMR prior, although not enough to reject the SHMR model outright.

As a summary of these results,
Figure~\ref{fig:mass_post} shows the distribution of stellar and dark mass within a three-dimensional aperture of 10~kpc, as well as the halo concentration for these two models.
For the SHMR-prior model, the data prefer a lower enclosed dark mass than in the prior, with a shift in the median $M_\mathrm{DM}$ within 10 kpc from $1.2\E{10} \ M_\odot$ to $5.1\E{9} \ M_\odot$.  
For the weak-priors model, the posterior distribution of $M_\mathrm{DM}(< \mathrm{10 \, kpc})$ extends all the way to the lower prior boundary ($\sim 10^2 \ M_\odot$), with a 90$^{\rm th}$-percentile upper bound of $1.2\E{8} \ M_\odot$.
Thus the data prefer a relatively low amount of DM within the region probed.

The more tightly constrained quantity of interest is the {\it total} dynamical mass within 10~kpc,
which is $(2.2^{+0.9}_{-0.6})\times10^8 M_\odot$, or dynamical $M/L_V = 1.7^{+0.7}_{-0.5}$.
The latter value is remarkably coincident with the independent stellar population estimate for DF2 (Section~\ref{sec:data}). We conclude that data-driven dynamical modeling of DF2 allows for at most an extremely low-mass DM halo, and suggests that this UDG is comprised purely of stars.

\section{Tidal Effects}\label{sec:tidal}

The models considered in the previous sections were for an isolated
dwarf and neglected any influence from
the nearby massive elliptical galaxy, NGC~1052.  
In particular, infall of a satellite into a larger host
initiates a process of tidal stripping, first from the outer DM halo,
then from the central regions, followed by total disruption.
Tidal stripping and heating has been proposed as the dominant
mechanism for forming UDGs, which could be considered as exemplars
for galaxies undergoing tidal disruption
\citep{Carleton2018}.  Some previously studied UDGs are clearly in the process of disruption \citep{Merritt2016}, while many others have undisturbed morphologies out to $\sim 4 \ \Reff{}$ \citep{Mowla2017}.

vD+18a presented analysis of tidal stripping to constrain the physical separation between DF2 and NGC~1052.  Here our aim is to develop a holistic model where the inferred UDG mass distribution is 
checked for consistency with tidal constraints, propagating uncertainties on viewing angle, satellite mass distribution, and central galaxy mass.
In particular, is a no-DM scenario implausible owing to  a likelihood of disruption?

\begin{figure}
    \includegraphics[width=\columnwidth]{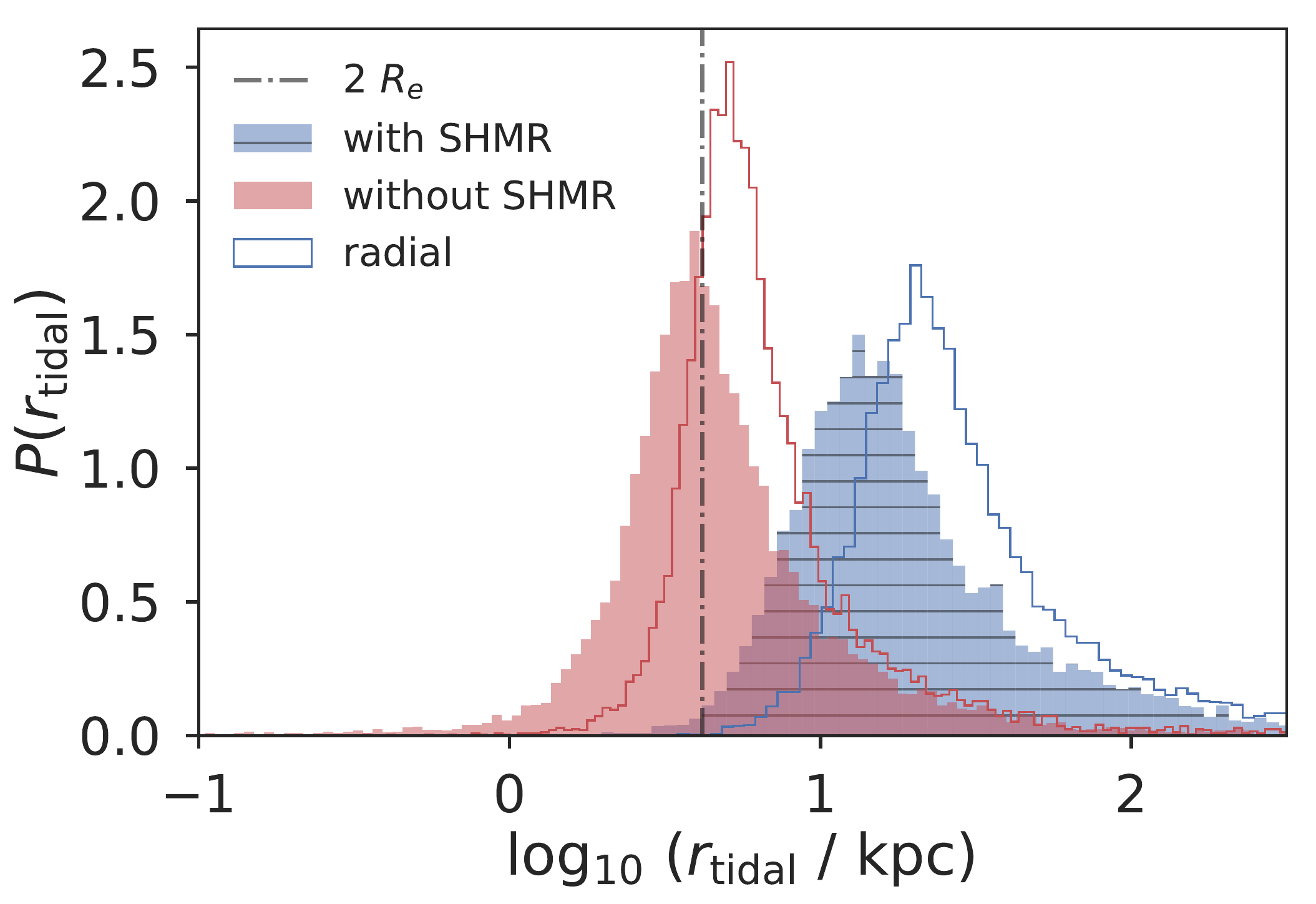}
    \caption{
    The distribution of DF2 tidal radii inferred for each of the two models.  The blue histograms show the limits inferred with a strong SHMR prior, while the red histograms show those for the model without the SHMR prior.  The filled histograms show the tidal radius from assuming a circular orbit, while the empty histograms show the same distributions from assuming a radial orbit.  
    The vertical dash-dotted line shows 2~$R_\mathrm{e}$ for the starlight.  
    We see that 52\% (81\%) of the no-SHMR-prior model samples for the circular (radial) orbit are above this lower bound, thus allowing for little/no-DM solutions that do not exhibit tidal disturbances.
    \explain{Changed text to clarify difference between radial and circular case}
    }
    \label{fig:tidal}
\end{figure}

We use a simple model for the tidal radius given enclosed masses of satellite and central galaxies:
\begin{equation}\label{eq:tidal}
	\rt = \left(\frac{M_\mathrm{sat}(\rt)}{(\alpha - \gamma_M) M_\mathrm{cen}(d)}\right)^{1/3} d ,
\end{equation}
where $d$ is the 3D distance between the two galaxies, $\gamma_M$ is the local log-slope of the enclosed mass profile of the central galaxy at $d$, and $\alpha = 2$ if we assume the orbit of the satellite is radial and $3$ if we assume the orbit is circular \citep{Bosch2018}.  \explain{changed text to clarify the meaning of $\alpha$}
\added{Without modeling any constraints on the actual orbit of DF2, we compare results assuming either radial or circular orbits for the satellite, assuming that the truth lies somewhere in between these two cases.}
For our sampled central mass profiles and separation distances, $\gamma_M \sim 1$.
DF2 shows no obvious evidence of tidal disturbances, with regular isophotes out to $\approx 2 R_{\rm e}$ ($\sim 4$~kpc; vD+18a).
This provides a tidal constraint 
that $\rt \gtrsim 4$~kpc.

To estimate the central-galaxy mass, we use the 
halo-to-stellar mass relation from \cite{Rodriguez-Puebla2017}.
%Eq.~66\footnote{We subtract 1 from their expression to match that of \cite{Behroozi2010}.}. 
For $M_* = 10^{11} M_\odot$ \citep{Forbes2017a}, we expect $M_\vir = 4.9 \times 10^{12} M_\odot$
with a scatter of 0.25~dex (from inverting the SHMR scatter of 0.15~dex).  We then adopt an NFW profile with concentration from the MCR and calculate the enclosed mass at a given radius.

To fold in all the uncertainties together (central mass, satellite mass posterior from the previous inference, and distance), we randomly sample from the underlying parameters, including a uniform distribution of projection angles.  We plot the resulting distribution of tidal radii in Figure~\ref{fig:tidal}.

For the data-driven model, $\rt = 4.3^{+4.7}_{-1.7}$ ($5.7^{+5.09}_{-1.71}$)~kpc when assuming a circular (radial) orbit. Thus there is a large fraction of model-posterior space (52\% for circular, 81\% for radial) where DF2 can have little/no DM yet be tidally undisturbed out to 4~kpc.
We note that the low-velocity star clusters observed out to $\sim$~7.5~kpc
could still be bound even with $\rt \sim 4$~kpc, 
if they have retrograde orbits
\citep{Read2006a}.

%\footnote{There is additional information in the observed line-of-sight velocity of DF2: $\sim 300 \kms$ relative to NGC~1052, which is comparable to the larger galaxy's circular velocity. If the satellite is bound rather than being a loosely-associated group member, then it is unlikely to be near apocenter, and the solutions with $\rt \gtrsim 100?$~kpc in Figure~\ref{fig:mean_rho} ({\it right}) can be excluded. We have experimented with including this constraint in our modeling using the orbit postulated in \citet{Ogiya2018}, but the results did not change substantially, while accurate inferences would require sampling a range of cosmologically motivated orbital parameters that is beyond the scope of this Letter.}

Turning to the SHMR-prior model, the dwarf would be naturally much more resistant to tides,
and the tidal radius would be farther out
(Figure~\ref{fig:tidal}).
However, the predicted value of $\rt = 16^{+28}_{-7}$~kpc (or $24^{+33}_{-10}$~kpc for the radial case) implies that DF2 would still likely have most of its DM stripped away by now, as $M_\mathrm{DM}(\rt) / M_\vir \sim 0.2$ ($\sim 0.4$). 

The latter point leads us to the possibility that DF2 started out with a normal DM halo,
but has been tidally eroded, not only by removing the outer parts but also by stripping 
out much of the central DM prior to disruption of the visible galaxy.
Such a solution was studied through $N$-body simulations by \citet{Ogiya2018},
who found that the final dark mass within 10~kpc could be $\sim 10^8 M_\odot$ --
which is consistent with our observations (see red curves in Figure~\ref{fig:mass_post}).
We note however two major caveats to this interpretation:
(1) \replaced{the very large initial DM core ($r_{\rm c} = r_s = 13$~kpc) in the Ogiya model may be implausible without non-standard DM physics}{there is a small range of orbital parameter space that allows for the necessary degree of stripping};
(2) the dynamical time within the UDG is comparable to its orbital
period, which may prevent it from relaxing into
a visually undisturbed system with cold kinematics.
\explain{We changed caveat 1 to be consistent with the accepted version of Ogiya 2018}

The difference in predicted tidal radii between the DM and no-DM models motivates looking beyond $4$~kpc for tidal features around DF2 to help distinguish between these two scenarios.

%tidal stripping by NGC~1052 could be responsible for a decrease in the amount of dark matter relative to the stellar mass \citep{Ogiya2018, Carleton2018}.  
%This process, along with other environmental effects, ought to change the SHMR we expect for satellite galaxies compared to that of central galaxies \citep[e.g.,][]{Rodriguez-Puebla2012, Yang2012, Reddick2013}.

%... caveat about assumption of dynamical equilibrium ...

%general idea that stripping should make GC overabundance problem worse not better (should be stripped preferentially first)

\section{DF2 in a Wider Context}\label{sec:context}

\begin{figure*}
	\includegraphics[width=\linewidth]{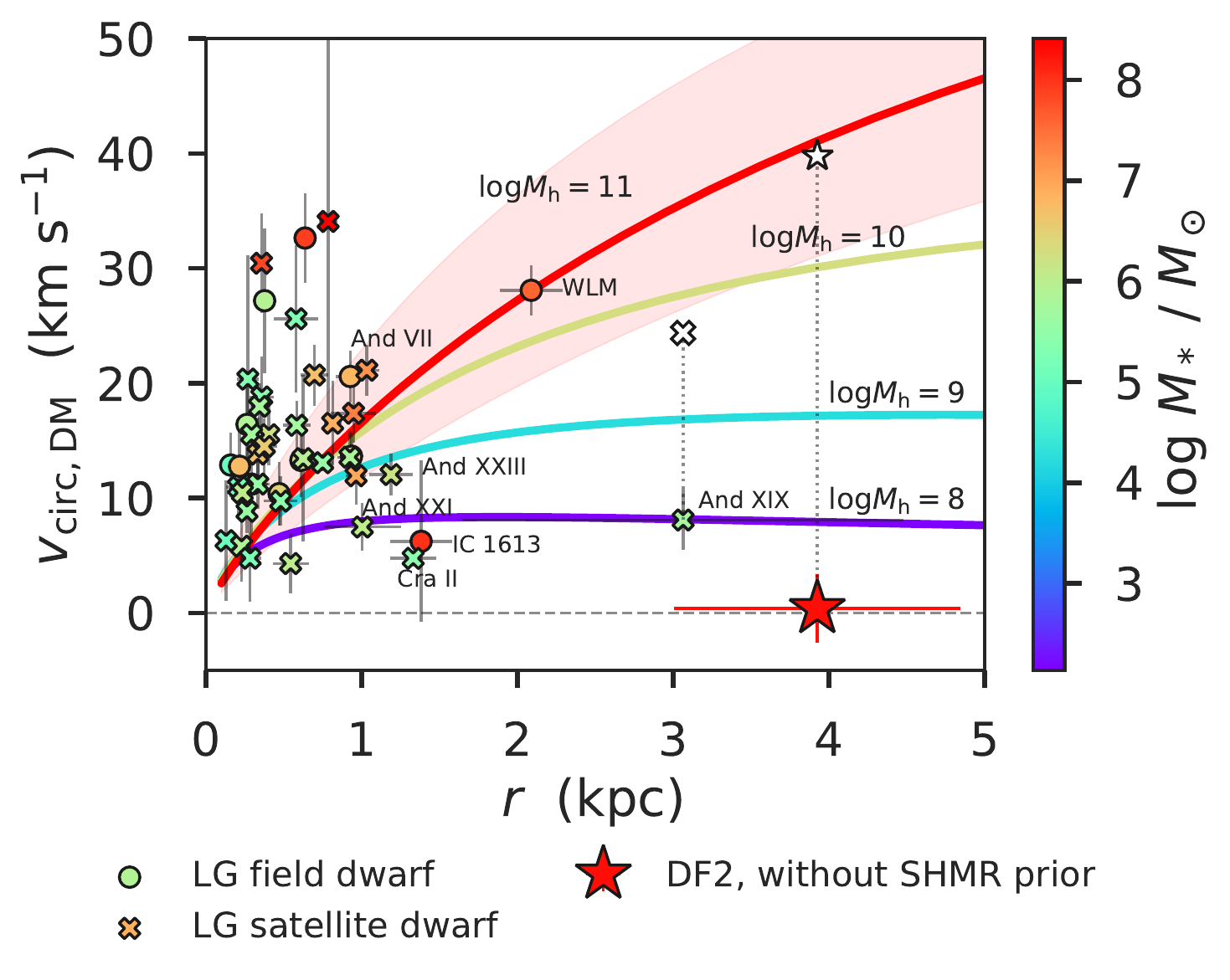}
    \caption{DF2 compared with Local Group dwarf galaxies.  The circles and x's show the circular velocity of the DM component for field and satellite dwarfs respectively.  The points are color-coded by stellar mass.  The curves show cored ($\gamma = 0.3$) NFW profiles for different halo masses (in $M_\odot$), color-coded by the mean expected 
    stellar mass.  The posterior predictive value for the data-driven DF2 inference is shown as the star, below $\log M_\vir / M_\odot = 8$.  The open markers with dotted lines for And XIX and DF2 show 
    the expected DM halos they would occupy given their stellar mass.  We see that DF2 is an outlier even beyond the extended LG dwarfs in both its size and in mismatch between expected and observed DM halo mass.}
    \label{fig:dwarfs}
\end{figure*}

We have found through Jeans modeling that the observations of cold kinematics in DF2 imply a very low DM content.  However, \citet{Martin2018a} disputed the unusual nature of this galaxy 
by noting its similar $\sigma$
and dynamical $M/L$ to previously studied dwarfs.
Here we emphasize that such comparisons neglect the different
measurement radii used, and we clarify the position of DF2 in a
wider context by constructing
a plot relating galaxy stellar masses, halo masses, and sizes (Fig.~\ref{fig:dwarfs}).

We take the compilation of Local Group (LG) dwarf galaxies from \cite{Fattahi2018}, selecting only galaxies with $M_* > 10^5 \ M_\odot$ and updating with sizes from \cite{Martin2016} where available.  Taking the dynamical mass within $r_{1/2} \approx 1.3 \Reff{}$ and subtracting the associated stellar mass, we compute the DM contribution to the circular velocity,
\begin{equation}
	v_\mathrm{circ, DM} = \sqrt{\frac{G M_\mathrm{DM}(< r_{1/2})}{r_{1/2}}} \ ,
\end{equation}
propagating uncertainties in the distance, size, luminosity, stellar $M/L$, and velocity dispersion.  We color these points in Figure~\ref{fig:dwarfs} by their stellar mass, with different symbols for field dwarfs versus satellites.
We compare these measurements with halo circular velocity profiles for several halo masses, adopting MCR concentrations and $\gamma = 0.3$ cores, while color-coding these profiles by the SHMR-predicted stellar mass.  The halo-concentration scatter is illustrated by the red band for the $10^{11} \ M_\odot$ halo.

This Figure shows that some dwarfs track cored-halo
profiles appropriate to their stellar masses.
Others have higher velocities and perhaps cuspy halos (e.g., \citealt{Spekkens2005,Onorbe2015,Genina2018}).
A few have low velocities; since most of these are satellites, 
they may be examples of ongoing tidal stripping that has
depleted their central DM content
(\citealt{Collins2013,Fattahi2018,Buck2018}).
DF2, however, stands out from all these galaxies by
having the lowest DM-velocity estimate, despite the much
larger measurement 
radius.
Andromeda~XIX is closest in $\sigma$--$r_{1/2}$ space but has $\sim 300\times$ lower stellar mass and
thus does not appear as DM-depleted as DF2.
IC~1613 has a high stellar mass but the smaller measurement radius allows for a larger range of halo masses.

We therefore strengthen the conclusion of vD+18a that DF2 is an 
extreme outlier in the usual dwarf--DM scaling relations.
There are then two main possible explanations.
One is that the galaxy formed with little or no DM, 
and the other is that it has been severely stripped of DM.
We cannot definitively discriminate between the two scenarios,
but in Section~\ref{sec:tidal} we pointed out potentially
major flaws in the tidal argument.
Furthermore, there is an additional clue that has so far been generally overlooked:
the very star-cluster system used to probe the dynamics of DF2 
is itself so far unique in the known Universe.
The clusters are on average far more luminous than in other galaxies including
the Milky Way, and they are also unusually extended and elongated \citep{Dokkum2018a}.
The presence of either a normal or a stripped DM halo provides no explanation for this novel 
observation.
%\footnote{It has been claimed that these massive star-clusters imply a relatively high mass for DF2, to prevent their in-spiraling on short timescales through dynamical friction \citep{Nusser2018}. However, this work did not consider the galaxy's flat central density profile, which might lead to star-cluster ``stalling'' \citep[e.g.,][]{Goerdt2006}}. 
On the other hand, if DF2 formed through a rare pathway without DM (e.g., scenarios discussed in vD+18a), then it is more plausible that its star-cluster system would show unusual properties as well.

The peculiar case of DF2 demonstrates the rich yield of information that can be obtained through detailed observations of dwarfs
beyond the Local Group,
which will help challenge and refine our understanding of galaxy formation and of the nature of DM.
\vskip 2pt

\acknowledgements
We thank the anonymous referee for useful suggestions.  AW thanks Viraj Pandya for helpful discussions.
This work was supported by NSF grants AST-1616598 and  AST-1616710.  AJR is a Research Corporation for Science Advancement Cottrell Scholar.  

\clearpage 

\bibliographystyle{aasjournal}
 \newcommand{\noop}[1]{}

\end{document}